\documentstyle[psfig,prb,aps]{revtex} 

\begin{document} 
\addtolength{\topmargin}{-1cm}
\title{\bf Structure and Bonding in Small 
         Neutral Alkali-Halide Clusters
         } 
\author{Andr\'es Aguado}
\address{Departamento de F\'\i sica Te\'orica, Facultad de Ciencias,
Universidad de Valladolid, 47011 Valladolid, Spain}
\author{Andr\'es Ayuela}
\address{
Institut f\"ur Theoretische Physik, Technische Universit\"at Dresden, 01062 Dresden, Germany 
}
\author{Jos\'e M. L\'opez and Julio A. Alonso}
\address{Departamento de F\'\i sica Te\'orica, Facultad de Ciencias,
Universidad de Valladolid, 47011 Valladolid, Spain}
\date{September 9, 1997}
\maketitle
\begin{abstract}
The structural and bonding
properties of
small neutral alkali-halide clusters, $(AX)_n$
with $n\le 10$, $A=Li^+, Na^+, K^+, Rb^+$ and $X=F^-, Cl^-, Br^-, I^-$, 
are studied using
the {\em ab initio} Perturbed Ion (PI) model
and a restricted structural relaxation criterion.
A trend of competition between rock-salt and hexagonal ring-like isomers is
found and discussed in terms of the relative ionic sizes.
The main conclusion is that an approximate value of $r_C/r_A=0.5$ (where $r_C$
and $r_A$ are the cationic and anionic radii) separates the hexagonal from
the rock-salt structures.
The classical electrostatic part of the total
energy at the equilibrium geometry is enough to explain these trends.
The magic numbers in the size range studied are $n=$ 4, 6 and 9,
and these are universal since they occur for all alkali-halides and
do not depend on the specific
ground state geometry. Instead those numbers allow for the formation of
compact clusters.
Full geometrical relaxations are considered
for $(LiF)_n$ ($n=3-7$) and $(AX)_3$ clusters, and the effect of Coulomb
correlation is studied in a few selected cases. These two effects
preserve the general conclusions achieved thus far.
\end {abstract}

\pacs{PACS numbers: 31.20.Gm 36.40.+d 61.46.+w}

\section{Introduction}
\label{sec:intro}

Small clusters often present
significant physical and chemical differences with respect to the bulk
phase.
While the structural possibilities are 
limited for the bulk material,
the number of different isomers which may coexist for clusters
is usually large, and the
energy differences between isomers are often small.
Here we are interested in neutral stoichiometric
clusters of typical materials with ionic bonding, that is $(AX)_n$ clusters, 
where $A$ is an alkali and
X a halide atom.
As {\em ab initio} studies on these clusters are computationally
expensive, the first theoretical calculations 
were based on
pairwise interaction models \cite{Mar83,Die85,Phi91}. 
Meanwhile, experimentalists moved
forward using several techniques to produce and investigate these clusters:
particle sputtering \cite{Cam81,Bar82}, where
rare-gas ions are used to bombard a crystal surface,
vapor condensation in an inert-gas atmosphere \cite{Ech81,Pfl85,Pfl86},
and laser vaporization of a crystal surface \cite{Con88,Twu90}.
In the expanding molecular
jet conditions, clusters undergo a rapid and 
efficient evaporative cooling, and
this leads to a cluster size distribution determined almost
exclusively by the cluster stability. 
The evaporative cooling process leads to the
so-called abundance magic numbers as a result of the longer time that the most
stable clusters remain in the beam before decaying. 
Alkali halide magic numbers are
often explained in terms of cuboid structures resembling fragments of
the crystalline lattice, but other possibilities like
ring stackings, or even mixed structures exist, which could be competitive.
Nevertheless, only
recently the possibility of
detecting different isomers has emerged in drift tube 
experiments \cite{Jar95,Mai96}. Those experiments have led to a renewed
interest in isomerization studies. 
In order to disentangle these
interesting problems, {\em ab initio} calculations \cite{Ayuel,Agu97,Wei92,Och94}
provide an ideal
complement 
to the experimental studies, which are restricted to charged
species. For instance,
a study of $(KCl)_n$
and $(LiF)_n$ clusters up to sizes of $n=32$ has been carried out
in ref. \onlinecite{Och94}, 
using
quantum-chemical
methods including correlation effects at the MP2 level.
We have performed
calculations for $(NaCl)_n$ and $(NaI)_n$
using the {\em ab initio} Perturbed Ion (PI) 
model \cite{Ayuel,Agu97}. 

The objective of the present paper is to give a global
characterization of the structure and other related 
properties of small neutral
alkali halide clusters. 
To this end we have carried out extensive PI calculations for the $(AX)_n$
$(n=1-10)$ clusters, identifying the most stable isomers, the binding
energy differences between some isomers, and the evolution of several
properties with the cluster size.
Trends are highlighted
and differences between different materials are discussed.
To obtain a more profound insight on the physics behind the observed trends,
the $(AX)_6$ clusters are studied in more detail.
Thus, conclusions can be drawn as to what energy
components dictate the structure of the ground state isomer.
Direct contact with the results of drift tube experiments is yet premature
since those experiments involve nonstoichiometric singly-charged clusters.

This paper is structured as follows: Section \ref{sec:theory}
contains a brief account of the theoretical method and computational details.
Section \ref{sec:results} contains the results for
the structure and bonding
properties.
Section \ref{sec:AX6}
contains a detailed study of the $(AX)_6$ clusters, which provides further
insight on the results obtained in section \ref{sec:results}.
Finally, section \ref{sec:conclusions} contains our conclusions and summarizes the principal
ideas of this study.

\section{Method}
\label{sec:theory}

According to the Theory of Electronic Separability (TES) \cite{Huz71,Lua87},
when a
system is composed of weakly interacting groups, its wave function
can be expressed as an antisymmetrized product of group wave functions.
If these satisfy strong-orthogonality conditions \cite{Lyk56,Par56}, the
total energy can be written as a sum of intragroup energies and 
intergroup interaction energies. 
The Perturbed Ion (PI) model is a particular application of the TES in which
each atom in
a cluster with a fixed structure (or in a crystal)
is considered as a different group \cite{Lua90}. 
The TES provides an efficient tool for dealing with ionic 
bonding \cite{McW94}. Thus, alkali-halide clusters are ideal systems to be 
treated by the PI model. The ions (positive and negative) are the basic 
entities of the model. Their electronic structures are selfconsistently
calculated, subject to the effect of the ion environment,
by using 
an effective hamiltonian including intragroup terms and coulombic, exchange
and projection ion-cluster interaction terms. 
The exchange interaction is accurately approximated by a nondiagonal spectral
resolution, as given by Huzinaga {\em et al.} \cite{Huz87}. The projection
energy term enforces strong orthogonality conditions between the ionic wave
functions, increasing their kinetic energies and
providing the short-range repulsive forces necessary for the
stability of the cluster. All the two-center integrals involved in the
calculation of these three interaction terms are analytically determined by
using the algorithm of Silverstone and Moats for the expansion of
a function around a displaced center \cite{Pen91a}.
The PI model is formulated at a Hartree-Fock (HF) 
level and has been described in
full detail in other papers \cite{Ayuel,Agu97,Lua90}. Correlation can also be
included in an approximate way.

In the calculations
we have used
large multi-zeta Slater-Type Orbital (STO) basis sets, taken from Clementi
and Roetti \cite{Cle74}.
Basis sets optimized for the description of ions in vacuum
are, in principle, an available choice to
describe those ions in a cluster.
However these are
not necessarily the best basis sets, as shown
in reference \onlinecite{Ayuel}(c).
The sensitivity of the PI model
to the quality of the wave function tails
is well documented \cite{Lua90,McW94,Hoj78,Pen91}. Specifically,
the effective potential for each ion depends, among other factors,
on the overlaps with
wave functions of neighbor ions. 
Thus, it is of paramount importance to 
choose
the most appropriate basis set for the description of
each material.
We have performed exploratory studies for
$AX$ molecules and $(AX)_6$ clusters.
Specifically, we have used the basis sets of Clementi and
Roetti with the exponents optimized for the description of ions in vacuum,
and also with the exponents optimized for the neutral species.
The main difference between these basis sets is indeed in the tail zone.
This leads to four distinct possible basis sets for each alkali halide.
The basis set leading to the largest binding 
energy for each individual alkali-halide material (the results for $(AX)_6$
and $AX$ lead to the same conclusions) has been
adopted for all clusters of 
that material.
We have also checked that
inclusion of diffuse orbitals
is not necessary.

\section {Cluster geometries and relative stabilities.}
\label{sec:results}

\subsection{Ground state structures and low-lying isomers}
\label{subsec:geometries}

The problem of minimizing the total cluster energy with respect to the 
positions of all the ions is computationally very demanding.
In our case we have performed a restricted search on
the (3n-6)-dimensional potential energy-surface.
The starting geometries have
been investigated by other authors \cite{Mar83,Die85,Phi91} within the context
of pair potential models. Specifically, we have 
considered cuboid structures (rock-salt fragments), 
ring-like configurations (mainly hexagonal) together with prismatic structures
obtained by stacking those rings, and some mixed configurations.
For the cuboid-like structures, the energy has been minimized with respect to
a single parameter, the nearest neighbor distance.
For ring-like structures,
we have relaxed two or three parameters, one accounting for the stacking
distance between parallel rings and the two 
others for the different distances of
cations and anions to the center of the ring. 

Figure 1 shows the results for $(LiF)_n$ and
$(KCl)_n$ clusters. The ground state and one or more low-lying isomers are
given for each $n$.
The energy difference
with respect to the most stable isomer is given (in eV) below each
isomer.
The first number corresponds to $(KCl)_n$ and the number below to $(LiF)_n$.
Although we have performed calculations for many alkali-halides, we only 
represent in Fig. 1 the results for $(KCl)_n$ and $(LiF)_n$ because those two
systems show well the main trend in structural stability, namely the
competition between rings and stacks of rings (mainly hexagonal) on one hand,
and structures that are fragments of a rock-salt crystal lattice on the other.
The tendency to form rings is stronger in $(LiF)_n$, and most $(KCl)_n$
clusters form instead rock-salt fragments. These structural trends support the
conclusion of Ochsenfeld and Ahlrichs \cite{Och94} of a slower convergence of
$(LiF)_n$ towards bulk properties. More generally, all alkali-halide clusters
containing $Li$ have a tendency to form rings, clusters containing $K$ and
$Rb$ form rock-salt pieces, and $Na$-halides represent an intermediate case.
Of course there are exceptions to this simple rule: for instance, we observe in
fig. 1 that the lowest energy structure of $(LiF)_7$ and $(KCl)_7$ is in both
cases a fragment of the wurtzite crystal. Also, the ground state structure for
$n=3$ is the hexagonal ring in both systems.

To give a more precise description of the competition between ring-like and
cuboid isomers we present in fig. 2 an structural stability map in which the
two coordinates are the empirical anion ($r_A$) and cation ($r_C$)
radii \cite{Ash76}. The map corresponds to $(AX)_6$, and the straight line
drawn in the map achieves a perfect separation between the systems in which the
lowest structure is the cuboid and those that prefer the hexagonal prism. The
same line separates the hexagonal prism and the rock-salt fragment in $(AX)_9$,
but the line may depend a little on $n$: a vertical line neatly separating
$Li$ clusters from the rest serves to distinguish between the octogonal ring
and the cube in $(AX)_4$. The conclusion from the structural maps is that two
parameters, $r_A$ and $r_C$, are enough to parametrize the competition between
ring-like and rock-salt isomers: alkali-halide clusters with a small $r_C$ and
a large $r_A$ have a tendency to form rings, although the first requirement
(small $r_C$) is almost sufficient.

The ring versus rock-salt competition can be further simplified to a 
one-parameter plot. In fig. 3 the energy difference between the two isomers
has been plotted versus the ratio $r_C/r_A$, again for $n=6$. A visible 
correlation exists between the two magnitudes. A critical ratio
$r_C/r_A=0.5$ separates the hexagonal from the
rock-salt structures.

\subsection{Evolution of the interionic distances with cluster size}
\label{subsec:distances}

In figure 4 we present the evolution of the averaged
interionic distances of
$(LiF)_n$ and $(KCl)_n$
with the cluster size for two isomeric families 
(rock-salt and hexagonal prism). 
The tendency is a slight increase of the cation-anion distance $d$ with cluster
size, but each isomeric family follows a different growth curve and
cation-anion distances are smaller in the hexagonal-ring pieces. Although not
plotted in the figure we have found that interionic distances in higher order
rings (octogonal, ...) are even smaller, so we conclude that the higher the
order of the rings forming the structure, the smaller the interionic distances.
$d$ tends to a saturation value 
in cubic $(KCl)_n$ clusters
which is about 0.1 \AA $~$
larger than the corresponding interionic bulk distance
$d(KCl)=3.33$ \AA \cite{Ash76}.
In cubic $(LiF)_n$ clusters, on the other hand, $d$ tends from below to a value
very close to the corresponding bulk limit $d(LiF)=2.02$ \AA $~$.

\subsection {Cluster Stabilities}
\label{subsec:stabilities}

Now we examine
the relative stability as a function of the cluster size.  The binding
energy per molecule of a given cluster $(AX)_n$ with respect to the separate
free ions is given by:
\begin{equation}
E_{bind} = \frac{1}{n} [nE_0 (X^-) +
nE_0 (A^+) - E(cluster)],
\end{equation}
where $E_0$ refers to the energies of the free ions.
In figure 5 we show $E_{bind}$ as a function of $n$, for a number of 
alkali-halides. The general trend is an increase of $E_{bind}$ with $n$.
However, some values of $n$ for
which the cluster is specially stable
can be observed. These are
$n=(4)$, $6$ and $9$. Local maxima can be seen for $n=6$, $9$ in all cases, 
and a
maximum or a change of the slope of the curve for $n=4$.
The most important feature is that these magic numbers are ``universal''
within the alkali-halide family, that is, they occur both in ring-forming
systems and in rock-salt forming systems and this occurs because the
difference in energy between ring-like and rock-salt isomers is small compared
to the change in binding energy when the cluster size $2n$ changes. In
summary, it is the especial value of $n$ that makes some clusters special
and not their particular ground state geometries. The stability occurs
because those special sizes permit the formation of ``compact'' clusters.
Let us illustrate this with specific examples. The two isomers of $(AX)_5$ 
(a decagonal ring and a cube with an $AX$ molecule attached to it) and
the hexagonal isomers 
of $(AX)_7$ contain some low coordinated ions, in contrast
to $(AX)_6$. $(AX)_9$ is also more
compact than the elongated forms of $(AX)_{10}$ and $(AX)_8$. The octogonal
prism in $(AX)_8$ contains also less coordinated ions than $(AX)_9$.
This idea of stability
of compact clusters is evidently associated to the
optimization of the attractive part of the electrostatic energy.
Excluding the lithium clusters, our results are in 
accordance with a geometrical model \cite{Mar83,Die85}
proposed to explain the magic numbers of large clusters with ionic bonding.
This model assumes that the most stable configurations correspond to those 
values of $n$ for which it is possible to form compact cuboid structures of type
$(a$ x $b$ x $c)$, where $a$, $b$ and $c$ are the number of atoms along 
three perpendicular edges.

\subsection{Inclusion of general geometrical distortions}
\label{subsec:distortions}

In addition to the restricted search of energy minima 
described in Section III-A above,
we have
performed ``full geometrical relaxations'' for
$(LiF)_n$ $(n\le 7)$ and for all the
$(AX)_3$ clusters. To this end, we have used a simplex
downhill algorithm \cite{Nel65,Wil91}.
The input geometries for these additional calculations are those
obtained from the previous restricted calculations.
Results for $(LiF)_n$ ($n =3-7$) are presented
in figure 6. Appreciable distortions are observed in some low-lying isomers
but not on the ground state.
The effect of the distortions is to reduce the energy difference between the
first isomer and the ground state, although the relative ordering of these two
is not changed. The effect is largest
for $n=5$, where the energy difference 
between isomers is lowered
by 3.39 eV. For $n=4$, the ring remains as the lowest energy structure,
but the isomers are now nearly degenerate.
For other materials
we have performed such calculations
only for $n=3$, and the distortions are very small.
The general conclusion of this section is the same as in Section III-A, namely,
that 
only $(LiX)_n$ and some $(NaX)_n$ clusters have a marked 
tendency to adopt ringlike
structures.

\subsection{Inclusion of correlation effects}
\label{subsec:correlations}

Coulomb correlation can play a significant role when the HF energy
differences between isomers are small. 
We have studied the influence of correlation in some selected cases
by using the unrelaxed Coulomb-Hartree-Fock (uCHF) 
model proposed by Clementi \cite{Cle65,Cha89}. 
Within this model, the PI wave functions 
calculated at the HF level
are kept fixed and the correlation energy is
added as a (perturbative-like) correction.
The calculations have been carried out using the restricted search described in
section \ref{subsec:geometries}. As the electron-electron repulsion is lowered upon inclusion of
correlation, a contraction of the interionic distances is obtained in all
cases. This contraction is always larger in the more compact (rock-salt) isomer
compared to the ring-like isomers.
Results for the effect on the binding energies of $(LiX)_4$ clusters 
are presented in Table \ref{tab:ax:table1}. The inclusion of correlation
leads to higher binding energies and
results in a larger stabilization of the cube isomer. This reduces the energy
difference between the two isomers in $(LiCl)_4$, $(LiBr)_4$, $(LiI)_4$, and
it changes the order of the two isomers in $(LiF)_4$, giving a
result in accordance with those of 
references \onlinecite{Och94,Hei95}, where a cube was obtained as the most stable
$(LiF)_4$ isomer.
However, $(LiCl)_4$, $(LiBr)_4$ and $(LiI)_4$ remain as octogonal rings.
Thus, the stability map for $n=4$ only changes gently and its main
characteristics remain valid upon inclusion of correlation.

We have also performed uCHF
calculations for the planar $(KX)_3$ isomers, and the
results are given in Table \ref{tab:ax:table2}.
The most stable structure for $(KX)_3$ at the HF level is the hexagonal 
ring, the same 
obtained for $(NaCl)_3$ in references \onlinecite{Ayuel,Wei92} and for $(NaI)_3$ in
reference \onlinecite{Agu97}. The calculations of Ochsenfeld {\em et al.} \cite{Och94} 
predict that the
rectangular (or
double chain) structure becomes 
the most stable $(KCl)_3$ isomer upon inclusion
of correlation at the MP2 level.
We do not obtain the double chain as the
minimum energy isomer of the $(KX)_3$ clusters upon including correlation, 
but the
energy differences between the two isomers decrease a little. For instance,
the energy difference for $(KCl)_3$ changes from 0.05 eV
at the HF level to 0.04 eV at the uCHF level.
On the other hand, if we perform a uCHF calculation for the fixed geometries of
$(KX)_3$ isomers fully relaxed at the HF level, 
the above energy differences are
0.14 (HF) and 0.11 eV (uCHF), respectively.
Thus,
inclusion of correlation
and a full geometrical relaxation have opposite effects on the 
relative stability of the two
isomers.
The main conclusion
is that both isomers could coexist in the experiments, since their energies
are close.

From figure 3 it can be appreciated that the two
$(NaBr)_6$ isomers have nearly the
same energy. Inclusion of Coulomb correlation inverts the order of the
isomers only
in this case. Thus, the general features of the stability map
remain valid after including correlation. In summary,
inclusion of Coulomb correlation results in a gain in
binding energy which is larger for
the rock-salt isomers, but this effect becomes significant
only for those cases showing near degenerate isomers.

\section{Detailed study of the
$(AX)_6$ isomers.}
\label{sec:AX6}

\subsection{Hexagonal versus rock-salt isomer}
\label{subsec:competition}

In this section we deal with the specific case of $n=6$ and work
at the level
of restricted relaxation explained in section \ref{subsec:geometries}.
Our goal is to achieve a deeper understanding of
the stability map presented in fig. 2. To this end we have
analysed the factors giving rise to the
cluster binding energy.
In order to study the deformations 
on the electron density of the ions 
induced by the cluster environment
we have calculated the expectation value
$<r^2>_{nl}$ for all the geometrically-inequivalent ions in the cluster. That 
expectation value is taken over the outermost occupied ionic orbital
$\psi_{nl}$.
In Table \ref{tab:ax:table3}, the values
of $<r^2>_{nl}$ for $F^-$ and $I^-$ anions in vacuum
and in some
$(AX)_6$ clusters are compared. 
Ions in the
hexagonal prism are labelled with the
letter $r$ (ring). In the rock-salt cluster there 
are two nonequivalent sites, labelled
$c$ (corner) and $e$ (edge).
The orbital contraction is important.
For a fixed anion, the contraction
is the largest for the $Li$ halide. 
In contrast, the contraction of the cationic orbitals is negligible.

It can be shown \cite{Ayuel,Agu97,Lua90} that 
the binding energy of equation (1) can be written as:
\begin {equation}
nE_{bind} 
= \sum_R E_{bind}^R
= -\sum_R (E_{def}^R + \frac{1}{2} E_{int}^R),
\end {equation}
where the sum runs over all the ions in the cluster.
According to this equation we can separate the binding
energy into a sum of site contributions,
each one composed in turn of two terms.
$E_{def}^R$ accounts for the self-energy associated to the
deformation of the wave functions of the free ions by the cluster environment,
so it is related to the change of $<r^2>_{nl}$ and 
depends on the specific
site in the cluster.
$E_{int}^R$ is the interaction energy of the ion R with the rest of the cluster,
namely
$E_{int}^R = E_{class}^R + E_{nc}^R + E_X^R +
E_{overlap}^R$, where the different terms are respectively: the electrostatic 
interaction between the ion
$R$ and the other ions of the cluster,
considered as point charges, the correction to this classical energy due
to the finite extension of the wave functions, the exchange part of the
interaction energy,
and the overlap energy \cite{Fra92}. Thus, after performing the $R$-sum,
the binding energy can accordingly be
partitioned as:
\begin{equation}
nE_{bind} = E_{def} + \frac{1}{2}E_{class} + 
\frac{1}{2}(E_{nc} + E_X + E_{overlap})
= E_{def} + \frac{1}{2}E_{int}^{classical} + \frac{1}{2}E_{int}^{quantum}.
\end{equation}
The deformation term is always positive \cite{Ayuel,Agu97}, 
so it opposes binding.
The overlap contribution dominates $E_{int}^{quantum}$, so this term
is also positive in all cases. Finally $E_{int}^{classical}$, which is
the Madelung interaction energy between point-like charges, is negative and
stabilises the cluster. It is worth to remark that all these energy 
components are obtained in a global selfconsistent process and that this
partition of the binding energy is not strictly necessary; nevertheless, it
proves conceptually quite useful. 

Now, the difference in binding energy per molecule between 
the hexagonal and rock-salt $(AX)_6$
isomers can be analysed in those three components. 
That is made in Table \ref{tab:ax:table4}. The following trends can be appreciated:
(a) The deformation part always favors the hexagonal isomer.
(b) The interionic distances are smaller in the hexagonal isomer and 
therefore the overlap is larger. Consequently
the term involving overlap favors the rock-salt isomer.
(c) $E_{int}^{classical}$ already displays
the main feature of figure 3, that is, the distinction between the hexagonal and
the rock-salt fragment.
This is very interesting because
a classical Madelung interaction would favor
the cuboid isomer if the interionic distances were the same in both isomers. 
The Madelung energy becomes more negative for the hexagonal prism only if the
interionic distance $d(hex)$ is smaller than a critical fraction
$\alpha d(cube)$ of the interionic
distance in the cube isomer, with $\alpha$ a number slightly smaller than 1.
This occurs for those materials with $r_C/r_A
\le 0.5$. Although the stability map can then be justified in terms of 
classical energy components, the equilibrium interionic distance,
which is a key factor in this discussion, results
from a delicate balance
between all the energy components, classical and nonclassical.
Returning to the borderline case
$(NaBr)_6$, the inclusion of Coulomb correlation induces a contraction of the
interionic distances in both isomers, but this contraction is 
larger for the rock-salt isomer, and the
Madelung energy produces an inversion in the order of the two isomers.

\subsection{Additional geometrical relaxation of the cuboid isomer}
\label{subsec:relaxation}

A better description of the structure of the rock-salt isomer
is obtained by allowing inequivalent ions to relax
independently. We have then relaxed the geometry
with respect to four parameters, the
distances from the cluster center to the four
inequivalent ion types (corner and edge cations
and corner and edge anions). 
In all the $Li$ compounds and in $(NaI)_6$, cations tend to move inwards,
while
anions move outwards, producing slight deformations.
This situation is reversed for the rest of clusters,
in which cations move outwards and anions 
move inwards.
The deformations have an interesting effect on the effective ion size. To see
this effect we
have calculated again $<r^2>_{nl}$ for the outermost orbital
of each ion.
Compared to the results for rock-salt isomers of Table \ref{tab:ax:table3} we have now obtained
almost identical contractions for the electronic clouds of anions at 
inequivalent (corner, edge) sites, with the conclusion that anions at
inequivalent positions have, nevertheless, an identical size. The size changes,
however, with the nature of the alkali partner.

\section {Conclusions}
\label{sec:conclusions}

Using the perturbed ion (PI) model, we have calculated the most stable
structures of neutral alkali-halide clusters $(AX)_n$, 
with $n\le 10$, $A=Li^+, Na^+, K^+, Rb^+$ and $X=F^-, Cl^-, Br^-, I^-$.
With few exceptions,
the equilibrium geometries obtained are 
ringlike structures for $(LiX)_n$ and $(NaI)_n$
clusters, and rock-salt fragments
for the rest of materials; $(NaBr)_6$ is a borderline case. 
The competition between rock-salt and ring-like isomers has been studied 
in detail and we
have found that an stability map with two parameters (the cationic and
anionic ionic radii) is
able to separate the alkali-halides in two well-defined structural families.
A further simplification to a one-parameter ($r_C/r_A$) plot has been
possible, from which an approximate value of 0.5 has been extracted that
separates the two families.
The interionic distances show a smooth variation with the number of
molecules in the cluster when different structural families are considered
separately. The alkali-halide magic numbers have an universal character,
based on the fact that highly compact isomers can be built with a number of
molecules equal to 4, 6 and 9. 
Additional calculations allowing for full geometrical relaxations
have been performed in the case of $(LiF)_n$ ($n=3-7$) and
$(AX)_3$ clusters. $Li$-based clusters
show larger geometrical distortions than clusters non-containing lithium. 
The structures have been compared with those obtained from 
other theoretical calculations, and the overall agreement is good. 
Inclusion of correlation corrections has been considered only in those
cases with near-degenerate isomers.

The partition of the binding energy 
for the $(AX)_6$ clusters has shown that 
the key ingredients to understand the structural differences of figures 2 and 3 
are the cation-anion distances and the classical electrostatic
Madelung contribution.

To sum up, 
a simple picture emerges to explain the observed structural
trends: when the ratio $r_C/r_A \le 0.5$,
the interionic distances at equilibrium, determined by minimization of the 
{\em total} energy, are such that the classical Madelung interaction
between point-like ions
favors the hexagonal isomer. 
On the other hand, when $r_C/r_A > 0.5$, the interionic distances produce
Madelung energies favoring rock-salt structures.

$\;$

$\;$

$\;$

{\bf ACKNOWLEDGMENTS}: Work supported by DGES (PB95-0720-C02-01) and Junta de
Castilla y Le\'on (VA25/95). A. Aguado is supported by a predoctoral fellowship
from DGES.

\newpage

{\bf Captions of tables}

$\;$

{\bf Table I}. Effect of correlation on the binding energies
of $(LiX)_4$ clusters.
HF and uCHF binding energies per molecule are given (in eV)
for the cube and the octogonal ring. 

$\;$

{\bf Table II}. Effect of correlation on the binding energies
of planar $(KX)_3$ clusters.
HF and uCHF binding energies per molecule are given (in eV)
for the double chain and the hexagonal ring.

$\;$

{\bf Table III}. Values of $<r^2>_{nl}$ in a.u. for the outermost
occupied orbitals of $F^-$ and $I^-$ anions in vacuum and
in four representative $(AX)_6$
clusters. r: ring site; c: corner site; e: edge site (as defined in
the text).

$\;$

{\bf Table IV}. Difference in binding energy per molecule
between rock-salt and hexagonal
$(AX)_6$ isomers, together with their partition in deformation, quantum
and classical interaction terms, as defined in the text. 
A minus sign indicates that the corresponding quantity favors the hexagonal
isomer.
All quantities in eV.

\pagebreak

% Table I. 
%
%\begin{table}
%\begin {center}
%
%\begin {tabular} {|c|c|c|c|c|}
% $Material$ & $basis$ $set$ $1$ & $basis$ $set$ $2$ & $basis$ $set$ $3$ & $basis$ $set$ $4$ \\ 
% & $cation-anion$ & $cation-neutral$ & $neutral-anion$ & $neutral-neutral$ \\ \hline
% LiF & 55.29$^*$ & 54.28 & 53.88 & 53.77 \\
% LiCl & 41.61 & 42.20$^*$ & 34.01 & 33.77 \\
% LiBr & 39.35 & 39.75$^*$ & 32.79 & 33.14 \\
% LiI & 35.54$^*$ & 33.74 & 30.86 & 29.17 \\
% NaF & 49.12 & 48.46 & 49.50$^*$ & 48.41 \\
% NaCl & 38.99 & 39.24 & 39.05 & 39.26$^*$ \\
% NaBr & 37.06 & 37.41 & 37.14 & 37.47$^*$ \\
% KCl & 34.01$^*$ & 33.93 & 29.06 & 28.68 \\
% KBr & 31.52 & 32.87$^*$ & 25.52 & 26.88 \\
% KI & 30.15$^*$ & 28.46 & 24.46 & 23.43\\
%\end{tabular}
%\end {center}
%\end {table}
%
%\newpage

% Table II.

\begin {table}
\begin {center}

\bigskip

\begin {tabular} {|c|c|c|c|c|c|c|c|c|} \hline
 Material &
\multicolumn {2}{c|}{$(LiF)_4$} &
\multicolumn {2}{c|}{$(LiCl)_4$} &
\multicolumn {2}{c|}{$(LiBr)_4$} &
\multicolumn {2}{c|}{$(LiI)_4$} \\
\hline
 & HF & uCHF & HF & uCHF & HF & uCHF & HF & uCHF \\
\hline
 Cube & 8.99 & 9.67 & 6.88 & 7.49 & 6.47 & 7.30 & 5.74 & 6.53 \\
 Ring & 9.00 & 9.59 & 7.00 & 7.52 & 6.56 & 7.32 & 5.91 & 6.57 \\
\end {tabular}

\end {center}
\caption{}
\label{tab:ax:table1}
\end {table}

% Table III

\begin {table}
\begin {center}

\bigskip

\begin {tabular} {|c|c|c|c|c|c|c|c|c|} \hline
 Material &
\multicolumn {2}{c|}{$(KF)_3$} &
\multicolumn {2}{c|}{$(KCl)_3$} &
\multicolumn {2}{c|}{$(KBr)_3$} &
\multicolumn {2}{c|}{$(KI)_3$} \\
\hline
 & HF & uCHF & HF & uCHF & HF & uCHF & HF & uCHF \\
\hline
 Double-chain & 6.41 & 6.88 & 5.25 & 5.75 & 5.06 & 5.70 & 4.63 & 5.33 \\
 Ring & 6.47 & 6.92 & 5.30 & 5.79 & 5.10 & 5.72 & 4.67 & 5.35 \\
\end {tabular}
\end {center}
\caption{}
\label{tab:ax:table2}
\end {table}

%% Table IV
%
%\begin {table}
%\begin {center}
%
%\bigskip
%
%\begin {tabular} {|c|c|c|c|c|c|} \hline
% & $Site$ & $<r^2>_{nl}$ & & $Site$ & $<r^2>_{nl}$ \\
%\hline
% $Li^+:$ $gas$ $phase$ & & 0.445 & $Rb^+:$ $gas$ $phase$ & & 3.438 \\
% $Li^+:(LiF)_6$ & $r$ & 0.436 & $Rb^+:(RbF)_6$ & $r$ & 3.410 \\
% & $c$ & 0.437 & & $c$ & 3.411 \\
% & $e$ & 0.435 & & $e$ & 3.402 \\
% $Li^+:(LiI)_6$ & $r$ & 0.443 & $Rb^+:(RbI)_6$ & $r$ & 3.429 \\
% & $c$ & 0.443 & & $c$ & 3.430 \\
% & $e$ & 0.443 & & $e$ & 3.427 \\
%\end {tabular}
%\end {center}
%\end {table}
%
% Table V

\begin {table}
\begin {center}

\bigskip

\begin {tabular} {|c|c|c|c|c|c|} \hline
 & $Site$ & $<r^2>_{nl}$ & & $Site$ & $<r^2>_{nl}$ \\
\hline
 $F^-:$ $gas$ $phase$ & & 2.207 & $I^-:$ $gas$ $phase$ & & 8.621 \\
 $F^-:(LiF)_6$ & $r$ & 1.794 & $I^-:(LiI)_6$ & $r$ & 7.575 \\
 & $c$ & 1.806 & & $c$ & 7.604 \\
 & $e$ & 1.763 & & $e$ & 7.398 \\
 $F^-:(RbF)_6$ & $r$ & 1.899 & $I^-:(RbI)_6$ & $r$ & 7.779 \\
 & $c$ & 1.907 & & $c$ & 7.789 \\
 & $e$ & 1.867 & & $e$ & 7.645 \\
\end {tabular}
\end {center}
\caption{}
\label{tab:ax:table3}
\end {table}

% Table VI

\begin {table}
\begin {center}

\bigskip

\begin {tabular} {|c|c|c|c|c|} \hline
$Material$ & $\Delta E_{def}$ & $\frac{1}{2} \Delta E_{int}^{quantum}$ & 
$\frac{1}{2} \Delta E_{int}^{classical}$ & $\Delta E_{bind}$ \\
\hline
 $LiI$ &  -0.069 & 0.048 & -0.089 & -0.110 \\
 $LiBr$ & -0.058 & 0.048 & -0.081 & -0.091 \\
 $LiCl$ & -0.061 & 0.059 & -0.088 & -0.090 \\
 $LiF$ &  -0.067 & 0.098 & -0.103 & -0.072 \\
 $NaI$ &  -0.034 & 0.055 & -0.041 & -0.020 \\
 $NaBr$ & -0.023 & 0.045 & -0.023 & -0.001 \\
 $NaCl$ & -0.031 & 0.033 &  0.003 &  0.005 \\
 $NaF$ &  -0.046 & 0.064 &  0.006 &  0.024 \\
 $KI$ &   -0.034 & 0.022 &  0.028 &  0.016 \\
 $KBr$ &  -0.022 & 0.013 &  0.029 &  0.020 \\
 $KCl$ &  -0.008 & 0.017 &  0.013 &  0.022 \\
 $KF$ &   -0.039 & 0.055 &  0.015 &  0.031 \\
 $RbI$ &  -0.034 & 0.014 &  0.044 &  0.024 \\
 $RbBr$ & -0.009 & 0.021 &  0.014 &  0.026 \\
 $RbCl$ & -0.001 & 0.020 &  0.008 &  0.027 \\
 $RbF$ &  -0.025 & 0.033 &  0.027 &  0.035 \\
\end {tabular}
\end {center}
\caption{}
\label{tab:ax:table4}
\end {table}

\pagebreak

\pagebreak

{\bf Captions of figures}

$\;$

{\bf Figure 1}. Lowest-energy structures and low-lying isomers of $(LiF)_n$ 
and $(KCl)_n$ clusters,
relaxed as indicated in the text. 
The energy
difference (in eV) with respect to the most stable structure is given below
the corresponding isomers. First row: $KCl$; second row: $LiF$.
Stability decreases from left to right 
for 
$(KCl)_n$ clusters. 

$\;$

{\bf Figure 2}. Structural stability map for $(AX)_6$. A plot in terms of the
cation and anion radii, $r_C$ and $r_A$ respectively, separates the
hexagonal (squares)
from the cubic (circles) isomers.
The same map is also valid for $(AX)_9$.

$\;$

{\bf Figure 3}. Energy difference between hexagonal 
and rock-salt isomers in $(AX)_6$ versus the
ratio of ionic radii.

$\;$

{\bf Figure 4}. Interionic distances in $(LiF)_n$ (open circles) and 
$(KCl)_n$ 
(full circles) clusters. Lines join isomers pertaining to the same
structural family: rock-salt (full line), hexagonal rings (dashed line).
Left scale is for $(LiF)_n$ and right scale for $(KCl)_n$.

$\;$

{\bf Figure 5}. Binding energy per molecule as a function of the cluster size
for some alkali-halide clusters. From the top to the bottom, these are:
$LiCl$, $LiBr$, $RbF$, $NaBr$, $LiI$, $KCl$, $KBr$, $RbCl$, $RbBr$, 
$KI$, and $RbI$.

$\;$

{\bf Figure 6}.
Lowest-energy structures and low-lying isomers of $(LiF)_n$ ($n=3-7$)
calculated allowing for a full geometrical relaxation.
Differences in total energy are given in eV.

\pagebreak

\pagebreak

\begin{figure}
\vspace{-10mm}
\psfig{figure=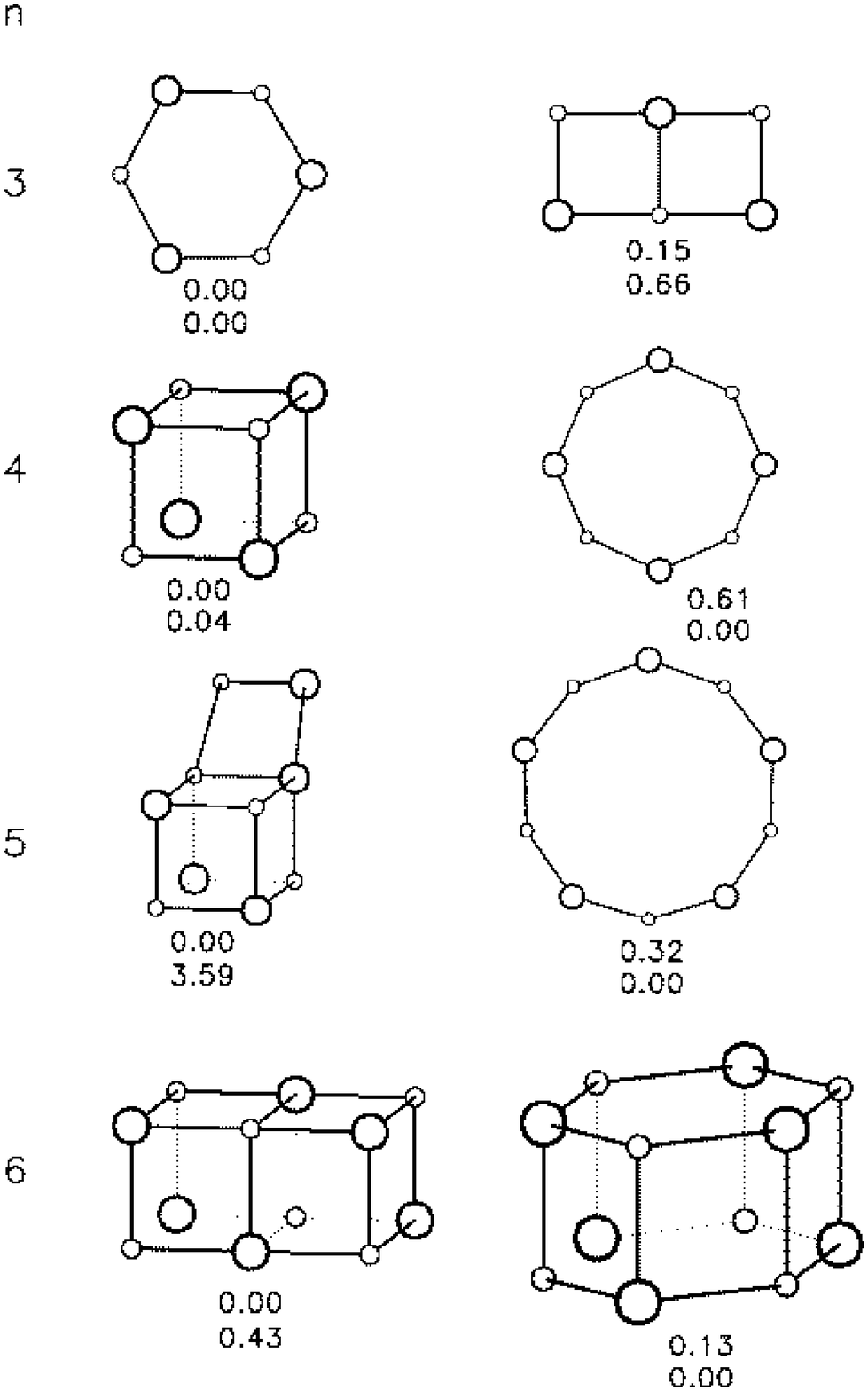,width=124.38mm,height=200.07mm,rheight=75mm}
\label{fig:ax:isomers}
\end{figure}
\pagebreak

\begin{figure}
\vspace{-10mm}
\psfig{figure=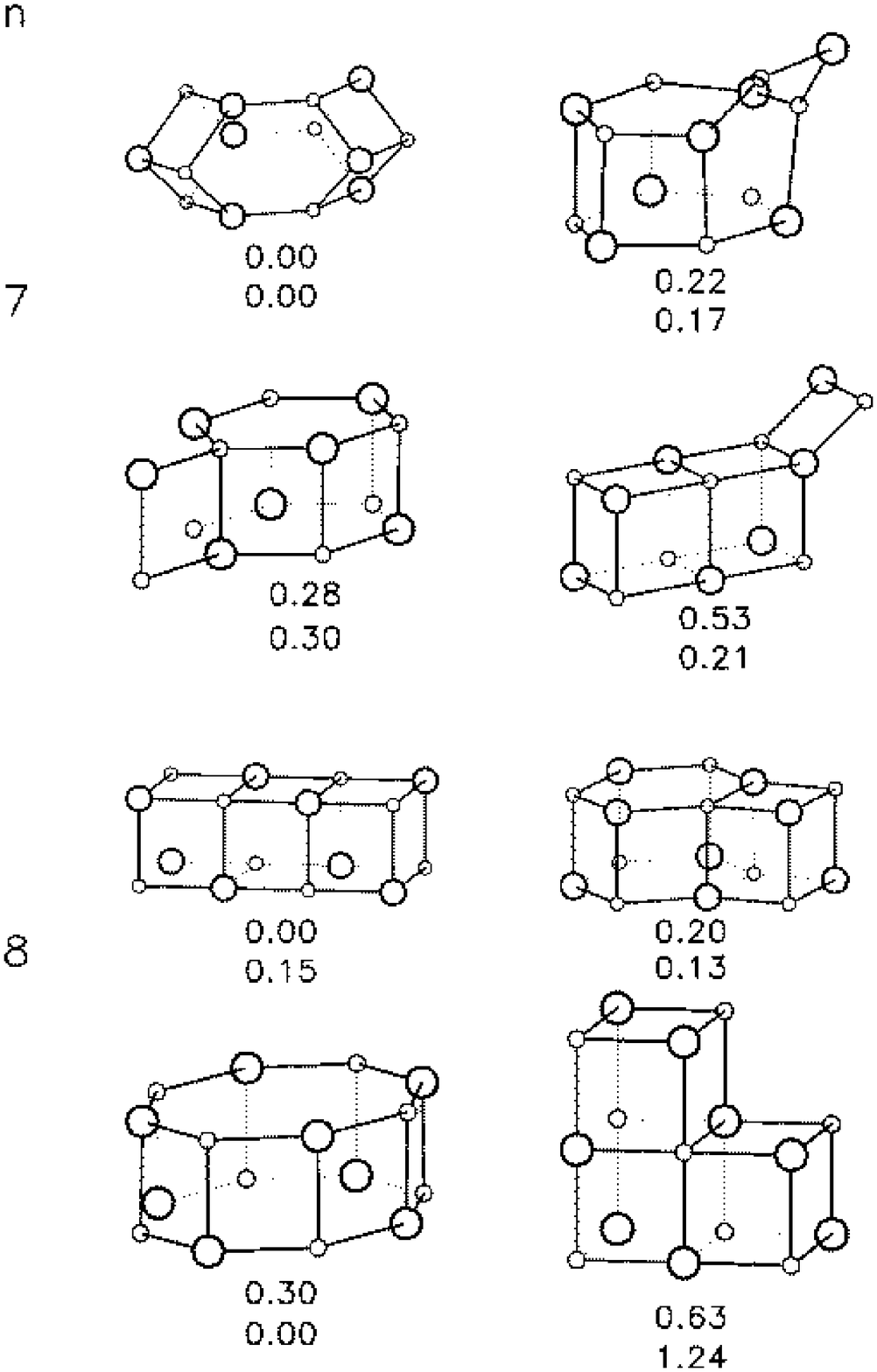,width=133.69mm,height=210.06mm,rheight=75mm}
\end{figure}
\pagebreak

\begin{figure}
\vspace{-10mm}
\psfig{figure=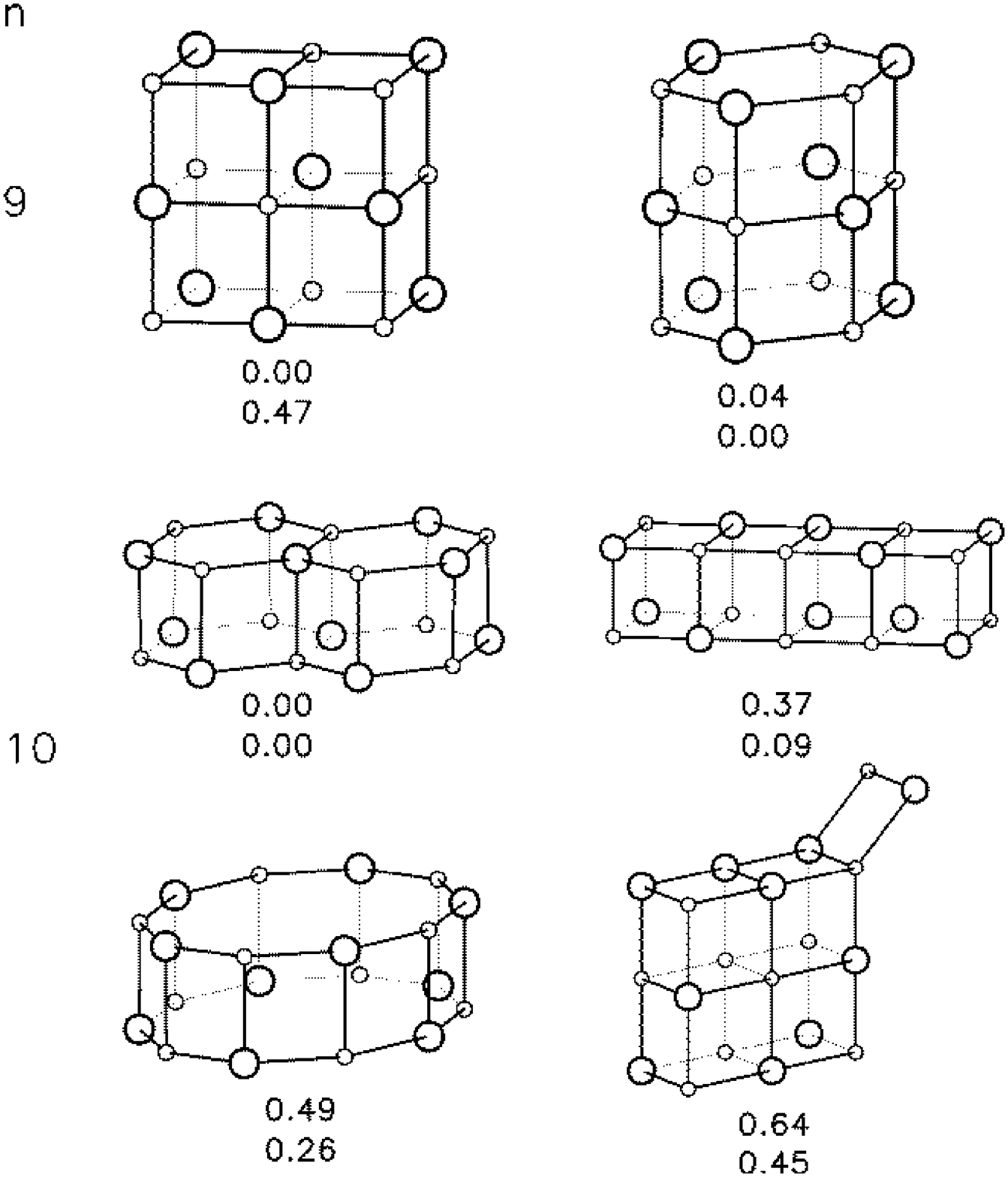,width=140.12mm,height=163.91mm,rheight=75mm}
\end{figure}
\pagebreak

\begin{figure}
\vspace{-10mm}
\psfig{figure=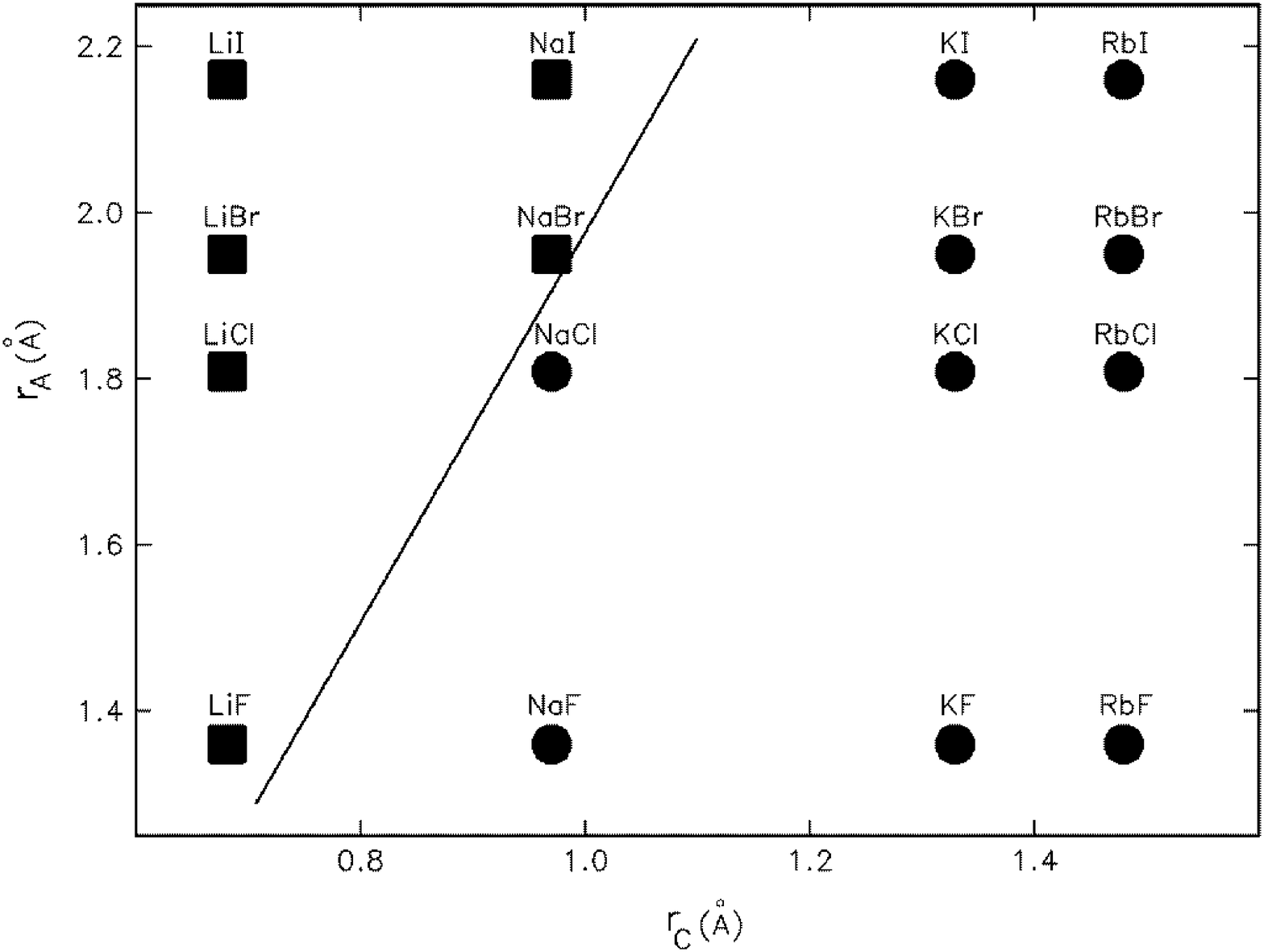,width=165.10mm,height=124.29mm,rheight=75mm}
\label{fig:ax:fig2}
\end{figure}
\pagebreak

\begin{figure}
\vspace{-10mm}
\psfig{figure=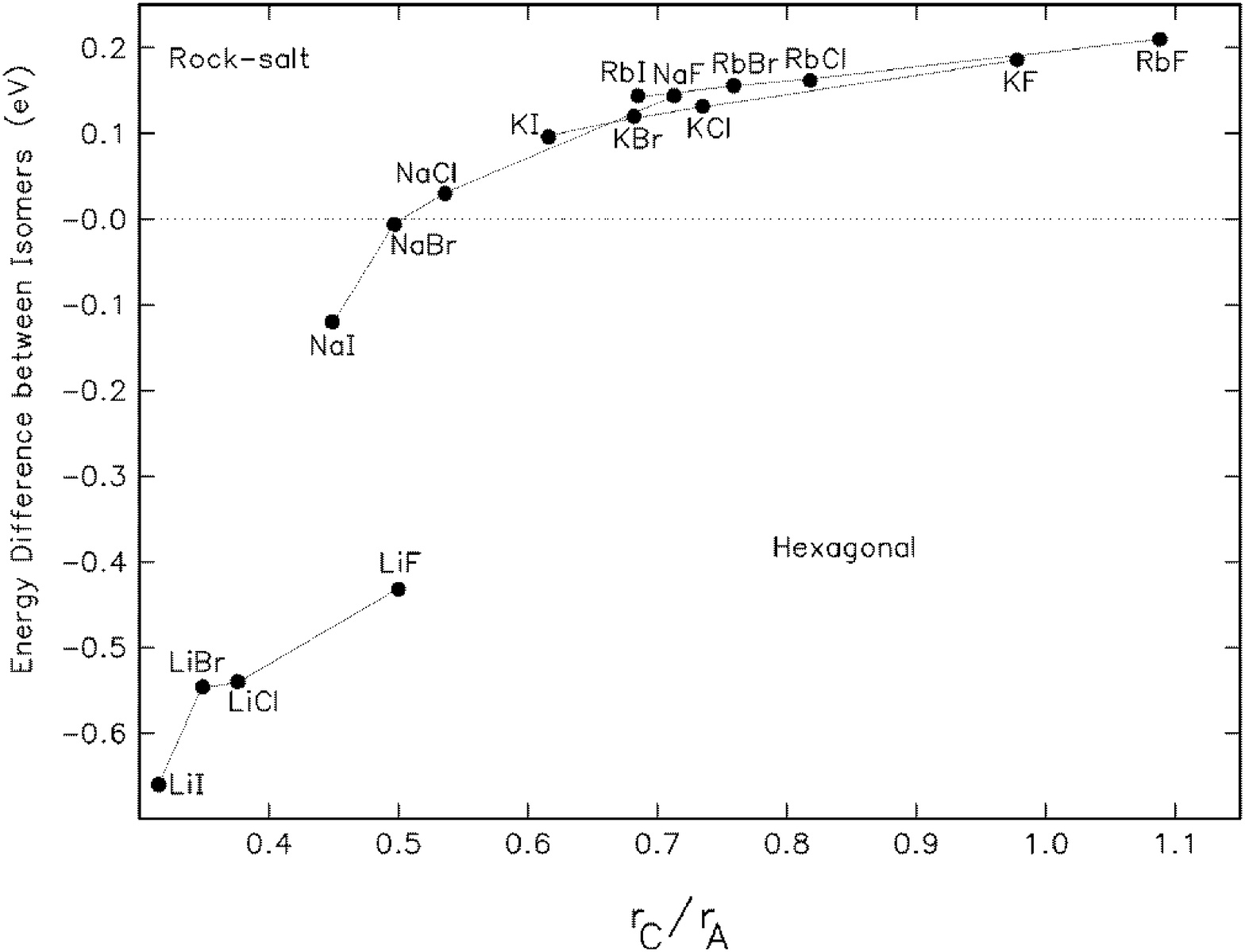,width=165.10mm,height=126.24mm,rheight=75mm}
\label{fig:ax:fig3}
\end{figure}
\pagebreak

\begin{figure}
\vspace{-10mm}
\psfig{figure=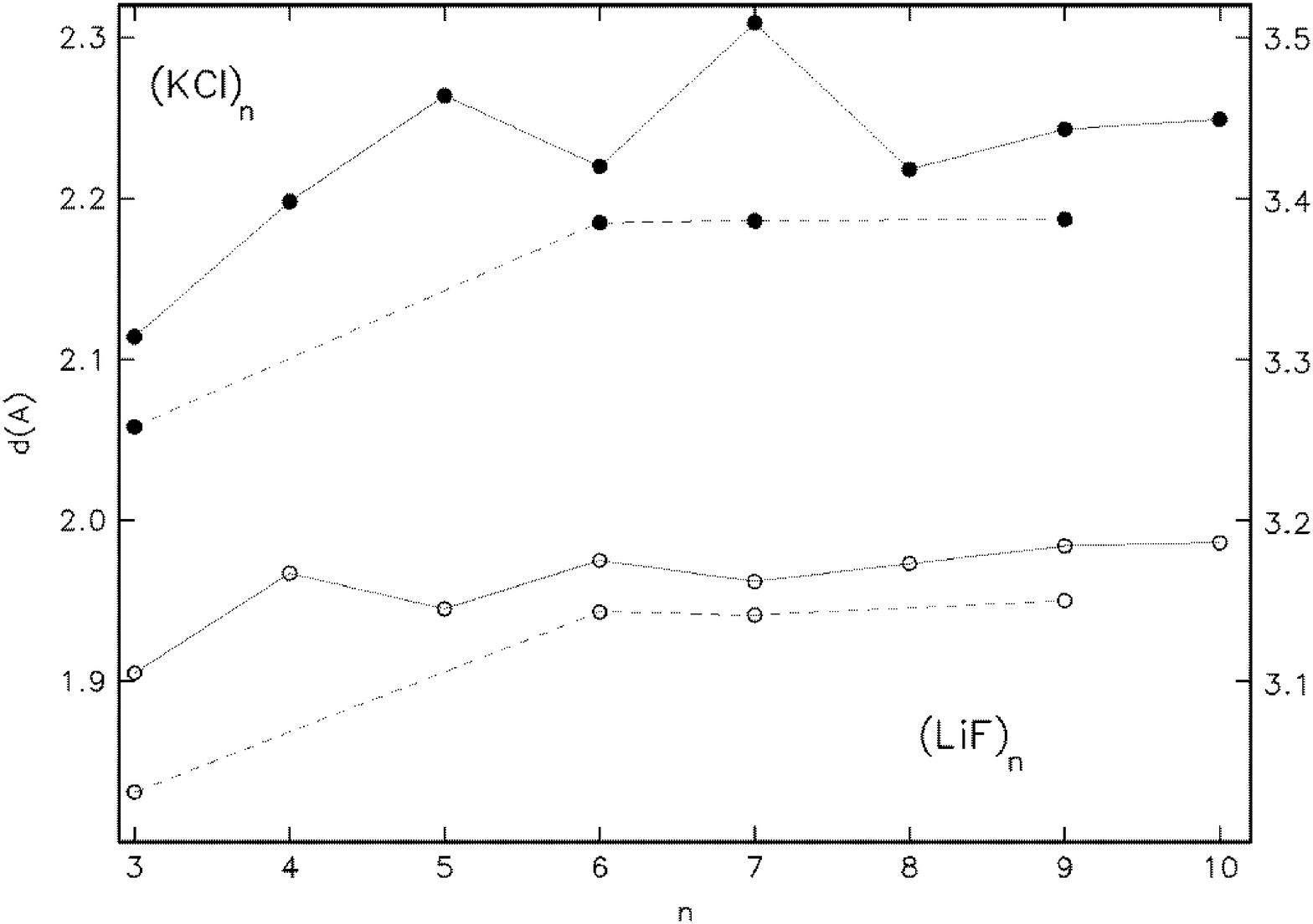,width=165.10mm,height=115.99mm,rheight=75mm}
\label{fig:ax:dist}
\end{figure}
\pagebreak

\begin{figure}
\vspace{-10mm}
\psfig{figure=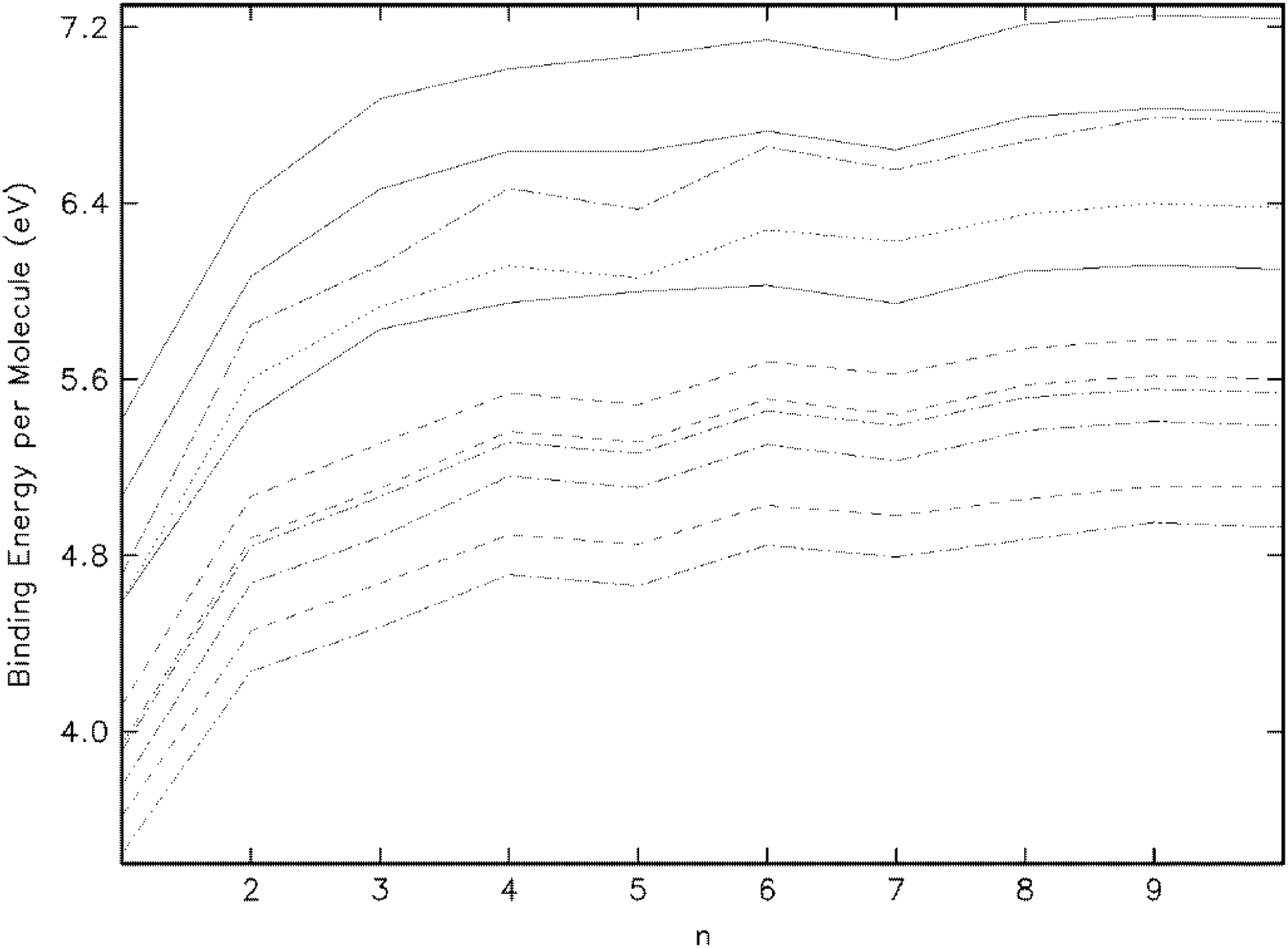,width=160.10mm,height=117.77mm,rheight=75mm}
\label{fig:ax:ebind}
\end{figure}
\pagebreak

\begin{figure}
\vspace{-10mm}
\psfig{figure=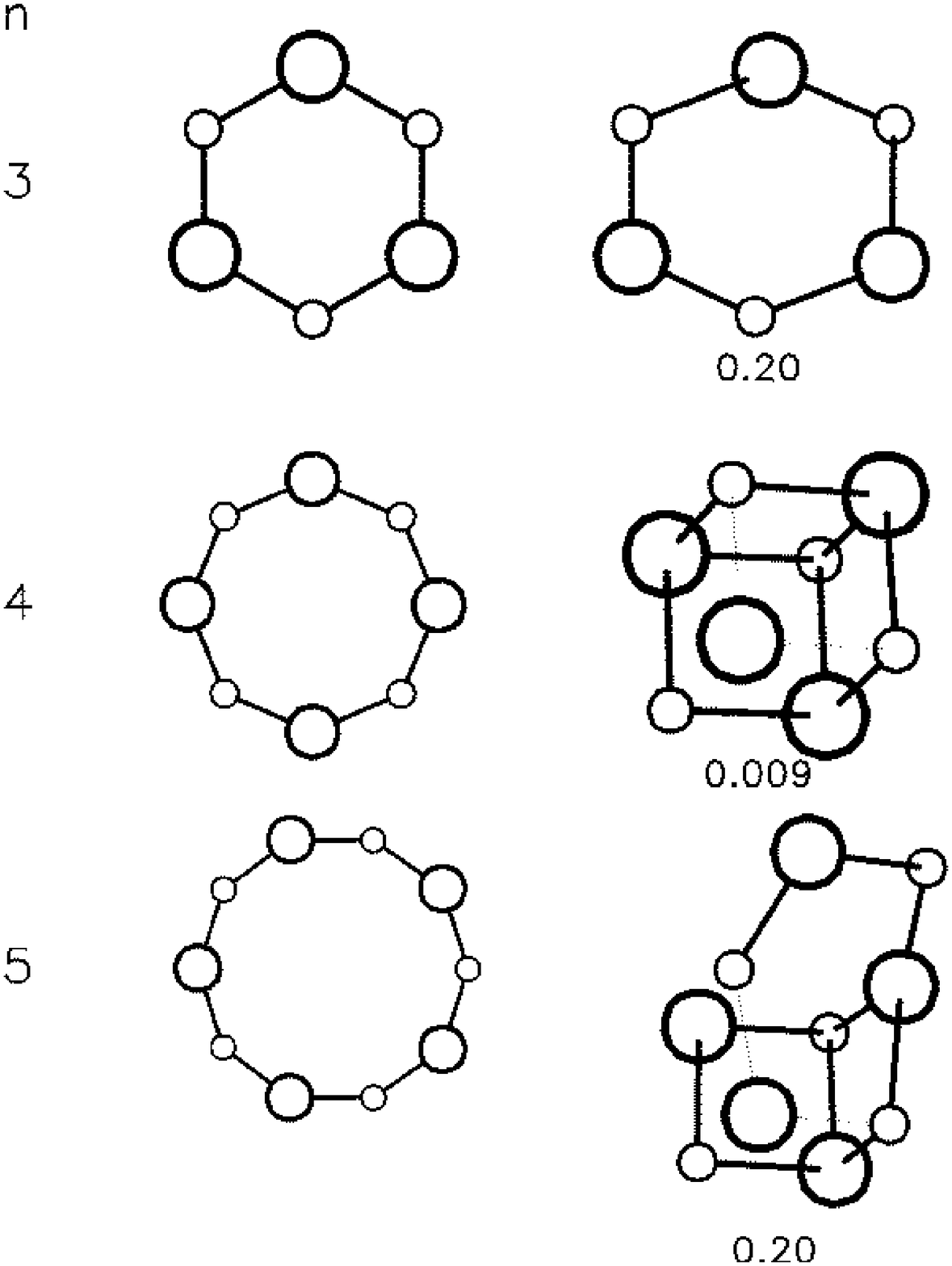,width=149.78mm,height=200.07mm,rheight=75mm}
\label{fig:ax:lifr}
\end{figure}
\pagebreak

\begin{figure}
\vspace{-10mm}
\psfig{figure=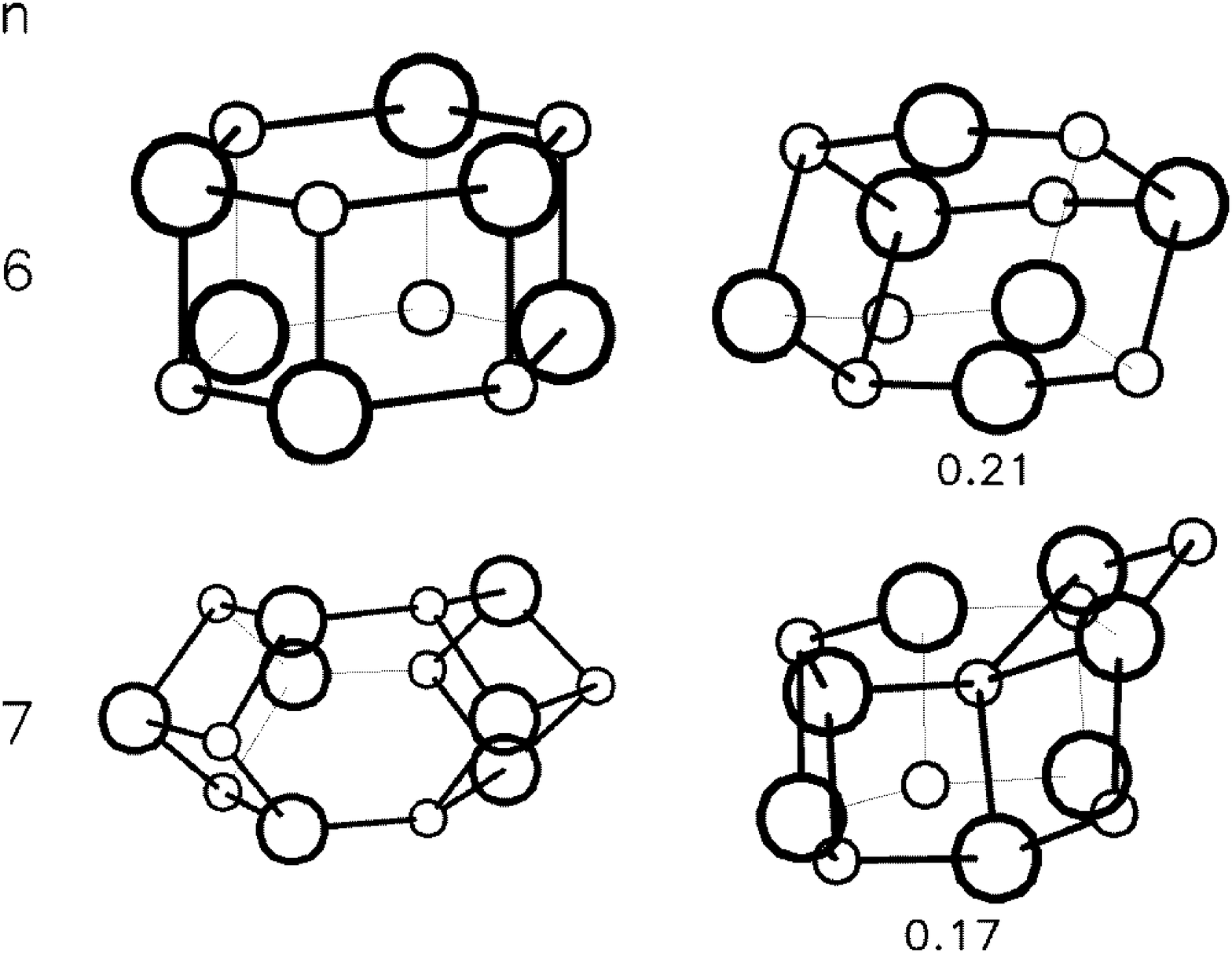,width=160.10mm,height=123.61mm,rheight=75mm}
\end{figure}
\pagebreak

\end{document}